# Electron-phonon coupling in $APd_3O_4$: A = Ca, Sr, and $Sr_{0.85}Li_{0.15}$

Bommareddy Poojitha, B. H. Reddy, Aprajita Joshi, Ankit Kumar, Asif Ali,

R. S. Singh, and Surajit Saha*

*Department of Physics, Indian Institute of Science Education and Research, Bhopal 462066,*

*India*

**Abstract**

Here we have investigated the role of electron phonon coupling on the Raman spectrum of narrow bandgap semiconductors $APd_3O_4$ (A = Ca, Sr) and hole-doped system $Sr_{0.85}Li_{0.15}Pd_3O_4$. Four Raman active phonons are observed at room temperature for all three compounds as predicted by factor group analysis. The lowest energy phonon (~190/202 cm$^{-1}$) associated with Pd vibrations is observed to exhibit an asymmetric Fano-like lineshape in all the three compounds, indicating the presence of an interaction between the phonon and the electronic continuum. The origin of the electronic continuum states and electron-phonon coupling are discussed based on our laser power- and temperature-dependent Raman results. We have observed an enhanced strength of electron-phonon coupling in $Sr_{0.85}Li_{0.15}Pd_3O_4$ at low temperatures which can be attributed to the metallicity in this doped compound.

**Introduction**

The interplay between lattice and electronic degrees of freedom is the key factor for various exotic phenomena in condensed matter physics and material science, such as quantum effects in thermodynamics, superconductivity, phonon and electron transport, phonon-assisted light absorption, and temperature-dependent optical properties of materials [1-4]. The electron-phonon coupling (EPC) reinforces the temperature-dependence of electrical resistivity in metals, carrier mobility in semiconductors, and also play a crucial role in materials exhibiting charge density waves (CDWs) as well as in the areas of spintronics and quantum information [5, 6]. On the other hand, semiconductors are the leading materials for next generation technologies to solve challenges in sustainable energy and environmental remediation [7]. Semiconductors with narrower band gap are widely used for infrared radiation detection, optoelectronic and thermoelectric device fabrication [8]. It was realized that the inclusion of EPC was necessary to interpret spectral and transport properties of semiconductors accurately [9]. In a semimetallic or narrow band gap semiconducting system, formation of excitons under suitable conditions may lead to a new exotic ground state, called excitonic insulator (EI) in which the Coulomb interaction and EPC are driving mechanisms [10].

In this context, in general, the nonmagnetic Pd transition metal and its transition metal oxides (TMOs) are receiving enormous attention due to their various ground states with the EPC controlled peculiar properties, owing to their unique crystal structure. For instance, the ground state and properties of Pd transition metal, lattice thermal conductivity (LTC) of $PdCoO_2$ are well controlled by the EPC [11-13]. The complex palladate oxides with narrow bandgap having significant Pd-4d states around the Fermi level ($E_F$) may allow the system to undergo an insulator to metal transition (IMT) upon hole/electron doping via monovalent/trivalent ion substitution at A-site [14-21]. For example, it was realized in PdO [15], $PdPbO_2$ [16, 17] and $APd_3O_4$ (A = Ca, Sr) systems [18-21]. Among all, $APd_3O_4$ (A = Ca and Sr) systems are potential candidates to explore the vibrational properties, possibility of EPC and IMT due to their unique crystal structure ($NaPt_3O_4$-type whose ground state and physical properties are driven by EPC [22]) and narrow band gap semiconducting nature (with the electronic bandgap value of 0.25 and 0.22 eV for $CaPd_3O_4$ and $SrPd_3O_4$, respectively obtained from transport measurements [23]). On the other hand, it was theoretically predicted that $CaPd_3O_4$ is a potential candidate for being an EI [24] although experiments suggested that it is unlikely to be [18]. Hence, it is very important to understand phonons and their correlation with electronic degrees of freedom in these systems. $APd_3O_4$ (A = Ca and Sr) systems may also be suitable for thermoelectric applications such as switching and sensing devices due to their narrow bandgap [17, 21, 23]. Since phonons and/or EPC have intimate relationship with thermoelectric properties, IMT, EI etc. [10, 25-27], and to the best of our knowledge, there are no phonon studies on these systems, we consider worth exploring phonons, possibility of EPC and IMT in these systems.

The optical spectroscopy is a powerful technique for investigating electronic and vibrational properties of a variety of systems and has provided extensive information and insights into the properties of solids [28]. For example, the strong electron-phonon coupling manifests itself in the form of asymmetric line shape of phonon mode in Raman or infrared spectrum [29-31], which can be analyzed by using Breit-Wigner-Fano (BWF) interference model. Here, we present the synthesis, crystal structure, and first experimental results of Raman spectroscopic measurements of the ternary palladates $CaPd_3O_4$, $SrPd_3O_4$ and $Sr_{0.85}Li_{0.15}Pd_3O_4$. The lowest energy phonon of $E_g/T_{2g}$ symmetry shows the Fano resonance due to EPC. The semiconductor to metal transition is observed by hole doping via $Li^{+1}$ substitution at the place of Sr in $SrPd_3O_4$. The temperature-dependence of asymmetry parameter ($1/q$) in all the three systems is in coordination with their electronic properties and band structure. More importantly, our results suggest that the electron-phonon coupling is tunable with chemical substitution and temperature in these systems thus making them potential candidates for various applications.

**Experimental details**

Polycrystalline Alkali earth ternary palladates $APd_3O_4$ with A = Ca, Sr, and $Sr_{0.85}Li_{0.15}$ were prepared by solid-state reaction route. High purity chemicals of $CaCO_3$, $SrCO_3$, PdO, and $Li_2CO_3$ from Sigma Aldrich were used as starting materials. A stoichiometric mixture of precursors was calcined at 630°C for 18 hours. Sintering was done at 700°C for a few times with intermediate grindings and pelletizing each time. Temperature-dependent powder X-ray diffraction (PXRD) measurements were done by using PANalytical Empyrean x-ray diffractometer with Cu-Kα radiation of wavelength 1.5406 Å attached with Anton Paar TTK 450 heating stage. Raman spectra were collected in the backscattering configuration using a LabRAM HR Evolution Raman spectrometer attached to a 532 nm laser excitation source and Peltier cooled charge-coupled device (CCD) detector. HFS600E-PB4 Linkam stage was used for temperature-dependent Raman measurements.

The phonons have also been calculated using projector augmented wave method (PAW) implemented in the Quantum Espresso code [32]. The optimization for total energy, lattice parameter, and phonon calculation was done using energy cut-off 150 Ry and 4×4×4 Monkhorst-Pack grids. The cubic lattice was kept fixed and equilibrium lattice constant of $SrPd_3O_4$ has been obtained by minimizing the calculated total energy as a function of lattice constant. We have chosen pseudopotentials generated using Perdew-Burke-Ernzerhof (PBEsol) exchange correlation.

**Results and discussion**

Alkaline earth palladates such as $CaPd_3O_4$ and $SrPd_3O_4$ crystallize in $NaPt_3O_4$-type crystal structure. The unit cell structure is drawn using VESTA (Visualization for Electronic and Structural Analysis) software [33] and displayed in Figure 1a. In this structure, each divalent Pd ion is coordinated with four $O^{-2}$ ions located in $PdO_4$ square planes. A-site atom is connected to eight $O^{-2}$ ions forming $AO_8$ cubes. The $AO_8$ cubes are connected to each other by $PdO_4$ planes through their edges [22, 34, 35]. Powder X-ray diffraction (PXRD) data collected on as prepared polycrystalline samples of $APd_3O_4$ with A = Ca, Sr, and $Sr_{0.85}Li_{0.15}$ (will be represented now on as CP, SP, and SLP, respectively) are refined by Rietveld method and displayed in Figure 1b. It confirms that powder samples are stabilized in cubic structure with the $Pm\bar{3}n$ (No.223) crystal symmetry. Rietveld refined lattice parameters and bond lengths for

CP, SP, and SLP are listed in Table-I which are in agreement with the earlier reports [23, 35-37]. The effective A-site radius [38] for SP, SLP, and CP is 1.260 Å, 1.226 Å, and 1.120 Å, respectively, that explains the observed trend (decrease) in the lattice parameters. The Pd-O bond length is similar to that found in PdO crystal (Pd-O = 2.03 Å) where Pd ions lie at the centres of rectangles and oxygen ions at each corner [39]. The Pd ion is divalent in these compounds with the electronic configuration $4d^8$. The 4d energy levels split into ($d_{xz}$, $d_{yz}$), $d_{xy}$, $d_{z^2}$, and $d_{x^2-y^2}$ in increasing order of energy due to the square planar configuration (Schematic is given in Figure S1 in supplementary material) [40]. The highest energy orbital, $d_{x^2-y^2}$, is empty with a small gap thus making $CaPd_3O_4$ and $SrPd_3O_4$ semiconducting in nature [41]. It is important to note here that the electronic structure calculations based on generalized gradient approximations (GGA) do not produce band gap suggesting that the electron-electron correlation plays an important role in these systems. It is GGA+U calculations which produces small gap for U ~ 4.5 eV (See Figure S2 in supplementary material). Other calculations based on hybrid functionals [23] also produce similar gaps in these systems confirming the role of electron correlation. Due to the small gap in these systems thermally- and/or photo- excited carriers can couple to the phonon degree of freedoms exhibiting the electron-phonon coupling with varying strength due to the temperature and laser power variations [42, 43].

To further investigate the effect of temperature on crystal structure and lattice parameters, we have performed temperature-dependent x-ray diffraction measurements in the range of 90 – 400 K. Figure 1(c, d, and e) shows the x ray diffraction patterns collected at a few temperatures for CP, SP, and SLP, respectively. Symbols in figure 1(f) show the temperature dependent lattice parameters obtained from Rietveld refinement of diffraction patterns for all three compounds. Fit using thermal expansion equation [44] as shown by lines in the figure 1(f) suggests no structural phase transition or anomalies in the lattice parameters and all the three compounds exhibit positive thermal expansion in the investigated temperature range (90-400 K).

Figure 2 shows the Raman spectra of $APd_3O_4$ (A = Ca, Sr, and $Sr_{0.85}Li_{0.15}$) collected at room temperature. According to the factor group analysis, the cubic structure with the $Pm\bar{3}n$ symmetry leads to four Raman active and five infrared (IR) active phonon modes with irreducible representations $\Gamma_{Raman} = 2\,E_g + 2T_{2g}$ and $\Gamma_{IR} = 5\,T_{1u}$ [45]. Four Raman active modes including two '$E_g$' and two '$T_{2g}$' are observed in agreement with the group theory predictions. The experimentally observed phonon frequencies at room temperature and corresponding mode symmetries are compiled in Table II. Our phonon calculations at the Γ-point of Brillouin zone based on Quantum Espresso support the mode assignments (see Table S1 in supplementary material). The mode P0 ($E^1_g$) and P1 ($T^1_{2g}$) are related to *Pd* vibrations, P2 is defect-induced Raman active phonon, whereas P3 ($T^2_{2g}$) and P4 ($E^2_g$) modes are associated with *O* vibrations [45]. To note that the mode P0 in CaPd$_3$O$_4$ and SrPd$_3$O$_4$ is too weak, possibly due to very low Raman scattering cross-section and could be seen only at low temperatures. The phonon frequency of all the modes increases systematically as we move from SP → SLP → CP which can be attributed to the decrease in the lattice parameters (as seen from PXRD discussed above). The lineshape of the modes P2, P3, and P4 can be well described by a Lorentzian profile whereas the mode P1 exhibits an asymmetric lineshape on the low-energy side of the band in the Raman spectra of all three compounds. This asymmetry can be described by the Breit-Wigner-Fano (BWF) interference model [31] (see Figure S3 in supplementary material). According to this model, the interference of auto-ionized state with a continuum gives rise to characteristically asymmetric peaks in excitation spectra. In low-gap semiconductors and metals, the asymmetry in a Raman lineshape arises from the interference between Raman scattering from the electronic continuum and that from a discrete optical phonon at the centre of Brillouin zone, provided both the Raman active excitations are coupled. The BWF model lineshape is given by [31,46]:

$$I = A\frac{\left(1+\frac{\epsilon}{q}\right)^2}{1+\epsilon^2}$$

$$= A\left[\frac{1}{q} + \frac{\left(1-\frac{1}{q^2}\right)}{1+\epsilon^2} + \frac{\left(\frac{2\epsilon}{q}\right)}{1+\epsilon^2}\right] \quad (1)$$

where $\epsilon = \frac{2(\omega-\omega_p)}{\Gamma}$ and $1/q$ is the degree of electron-phonon coupling that describes the departure of the lineshape from a symmetric Lorentzian function, *A* is the maximum intensity of the BWF spectra, $\omega_p$ and $\Gamma/2$ are the real and imaginary part of phonon self-energy, respectively. In Eq. 1, the first term $\left(\frac{1}{q}\right)$ represents a constant continuum spectrum, the second term is for a discrete Lorentzian spectrum while the interference effect between both spectra is accounted by the third term $\left(\frac{\left(\frac{2\epsilon}{q}\right)}{1+\epsilon^2}\right)$. The interference term resembles the ratio between the probability amplitude of the continuum spectrum to that of the discrete spectrum, giving rise to an asymmetric lineshape in the Raman spectrum. The asymmetric Raman lineshape (Eq. 1) takes a Lorentzian profile when there is no electron-phonon coupling i.e. when $\frac{1}{q} = 0$. Figure 3 shows the Raman spectra collected at a few different laser powers using 532 nm radiation. A clear redshift in the peak position and an increase in the asymmetry are observed with increasing laser power. The insets of Figure 3 display corresponding Stokes and anti-Stokes spectra collected at the highest laser power used (2.37 mW) which clearly shows that both the Stokes and anti-Stokes spectral shapes are asymmetric. The spectra are fitted with BWF lineshape and the corresponding parameters (frequency, linewidth, and asymmetry parameter) are plotted as a function of the laser power as shown in Figure 4(b, d, f). The phonon frequency (linewidth) decreases (increases) with increasing laser power while the asymmetry parameter ($1/q$) increases with increasing laser power. The increase in $\frac{1}{q}$ with laser power is photo-induced effect where photo-excited (and thermally excited due to laser heating at higher powers) carriers couple to the phonons. The corresponding parameters of the mode P1 in the anti-Stokes spectrum also show similar behaviour (Not shown here). The redshift in the phonon frequency can be attributed to the laser heating effect. The mode softening and symmetric (Lorentzian) broadening of the phonon modes P2, P3, and P4 can be attributed to the lattice anharmonic effects (see Figures S4 and S5 in supplementary material). In order to understand the temperature-dependence of the electron-phonon coupling, as indicated by the asymmetric lineshape of the mode P1, we have collected the Raman spectra in the temperature range of 80-600 K (Figure S5 in supplementary material). The temperature-dependence of the remaining phonon modes is given and discussed in supplementary material.

Figure 4 displays the profile for P1 mode at a few temperatures along with Fano fitting (solid lines in Figure 4(a,c,e)). The extracted phonon parameters (frequency and linewidth) are plotted as a function of temperature for all the three compounds in Figure 4(b,d,f). The mode shows clear redshift in frequency and systematic broadening of the linewidth with increasing temperature in the investigated temperature range (80-600 K). The temperature-dependence of frequency and linewidth are analysed by using anharmonic model [47, 48] (see Figures S7-S8 in supplementary material) as represented by the solid lines in Figure 4(b,d,f), showing the best-fit. Hence, the mode softening (and line broadening) with increasing temperature is attributed to the lattice anharmonicity. Figure 5 presents the inverse of BWF asymmetry parameter ($1/q$) which is directly related to the strength of electron-phonon coupling ($\lambda_{el-ph}$), as a function of temperature for all the three compounds. As can be seen from Figure 5 (a,b,c), the coupling is almost comparable in all the three palladates, however, showing a weak temperature-dependence and differences with each other. The value of *1/q* increases with increasing temperature in $CaPd_3O_4$, whereas it slowly decreases for $SrPd_3O_4$ and very clearly decreases with increasing temperature in $Sr_{0.85}Li_{0.15}Pd_3O_4$. It should be noted that the linewidth of P1 mode in all the three compounds shows a very similar behaviour with temperature irrespective of the small differences in the temperature-dependence of asymmetry parameter (*1/q*). Hence, the major contribution to the linewidth comes from phonon-phonon anharmonic interactions in these systems.

In order to further understand the electronic structure and IMT along with the origin of electron-phonon coupling, these compounds have also been investigated using photoemission spectroscopy which will be reported elsewhere [49]. The photoemission measurements suggest that $CaPd_3O_4$ is slightly more insulating than $SrPd_3O_4$ while $Sr_{0.85}Li_{0.15}Pd_3O_4$ is metallic in nature [49]. Based on this and our Raman data we propose that there are contributions of the thermally-and also photo-excited carriers in the observed signatures of electron-phonon coupling. An increase in the EPC (*1/q*) with increasing temperature in insulating CP, as shown in Figure 5a, indicates the role of thermally excited carriers getting coupled to the phonon. On the other hand, the decreasing trend of EPC (i.e. *1/q*) with temperature in SP and SLP (Figures 5b, c) clearly indicate toward their excited carrier-induced metallic nature. In metallic systems, as the temperature increases the EPC decreases, as reported for $LaMnO_{3+\delta}$ - shown in Figure 5d [42]. This can be attributed to an enhanced electron-electron scattering in metallic systems at high temperatures leading to reduced EPC. Further, we note that the strength of EPC (i.e. *1/q*) observed in the $APd_3O_4$ compounds are very much comparable to the well-known systems like (a) metallic carbon nanotubes (1/q: 0.04 to 0.10 depending on the diameter) [50], $LaMnO_{3+\delta}$ thin films (1/q: 0.2 to 0.4 varying with temperature and oxygen partial pressure during the growth) [42], Si nanowires (1/q: 0.06 to 0.12 depending on laser power) [43], and one-layer graphene [51] etc. as shown in Figure 5d. The reasonably high EPC in narrow band-gap $APd_3O_4$ compounds makes them potential candidates for various possible applications in electronics, opto-electronics, and optical metamaterials.

**Conclusions**

In summary, we have synthesized polycrystalline compounds of $APd_3O_4$ with A = Ca, Sr, and $Sr_{0.85}Li_{0.15}$ by conventional solid-state reaction route. Four Raman active modes are observed in agreement with the group theoretical predictions. The lowest-energy optical phonon associated with palladium vibrations shows the Fano asymmetry arising from the electron-phonon coupling in all the three compounds. The EPC can be controlled by the temperature and carrier doping in these systems. The difference in the temperature-dependence of the asymmetry parameter in these three compounds is attributed to the relative differences in their electronic structure. Interestingly, the basic ingredients required for superconductivity such as electron-electron correlation along with electron-phonon coupling are available in these narrow band gap semiconducting palladates. Thus, we believe that our work will motivate researchers to explore such possibility in electron/hole dopped systems.

**Acknowledgement**

Authors acknowledge IISER Bhopal for research facilities, B. P. acknowledges the University Grant Commission for fellowship and S. S. acknowledges SERB (project Nos. ECR/2016/001376 and CRG/2019/002668) and Nano-mission (Project No. SR/NM/NS-84/2016(C)) for research funding. Support from DST-FIST (Project No. SR/FST/PSI-195/2014(C)) is also thankfully acknowledged. Authors acknowledge the support of Central Instrumentation Facility and HPC Facility at IISER Bhopal.

**Availability of Data:** The data that supports the findings of this study are available within the article [and its supplementary material].

Table I. Room Temperature lattice parameters and bond lengths extracted from x-ray diffraction results

| Compound | Lattice parameter a (Å) | Bond lengths (Å) | | |
|---|---|---|---|---|
| | | Pd-O | A-O | A-Pd |
| $SrPd_3O_4$ | 5.8277 | 2.0581 | 2.5230 | 3.2530 |
| $Sr_{0.85}Li_{0.15}Pd_3O_4$ | 5.8194 | 2.0575 | 2.5199 | 3.2531 |
| $CaPd_3O_4$ | 5.7391 | 2.0290 | 2.4851 | 3.2082 |
| Calculated lattice parameter for $SrPd_3O_4$ is 5.82 Å | | | | |

Table II: Frequency of Raman active phonons at room temperature.

| | Space group $Pm\bar{3}n$ (No.223) and point group $O_h$ ($m\bar{3}m$) | | | | |
|---|---|---|---|---|---|
| | Wyckoff positions Sr/Ca/Li: 2a Pd: 6c O: 8e | | | Irreducible representations $T_{1u}$ $E_g + T_{2g} + 2T_{1u}$ $E_g + T_{2g} + 2T_{1u}$ | |
| | Total irreducible representations $\Gamma_{Raman} = 2\,E_g + 2T_{2g}$ and $\Gamma_{IR} = 5\,T_{1u}$ | | | | |
| **Mode** | **Symmetry** | **Frequency (cm$^{-1}$)** | | | **Atoms involved in vibration** |
| | | $SrPd_3O_4$ | | $Sr_{0.85}Li_{0.15}Pd_3O_4$ | $CaPd_3O_4$ | |
| | | Cal. | Expt. | Expt. | Expt. | |
| P0 | $E_g^1$ | 177 | 185* | Too weak to be observed | 200* | Pd |
| P1 | $T^1_{2g}$ | 193 | 190 | 193 | 202 | Pd |
| P2 | Defect-induced Raman-active | -- | 508 | 511 | 554 | O |
| P3 | $T^2_{2g}$ | 512 | 546 | 553 | 592 | O |
| P4 | $E_g^2$ | 539 | 553 | 556 | 599 | O |
| Calculated Acoustic/IR active phonon frequencies with $T_{1u}$ symmetry are 0 [Sr], 151 [Pd], 161 [Pd], 370 [O], and 551 [O] cm$^{-1}$. The atoms involved in corresponding vibration are mentioned in the square brackets. | | | | | | |
| * indicates the weak modes and frequency value is given at 80 K | | | | | | |

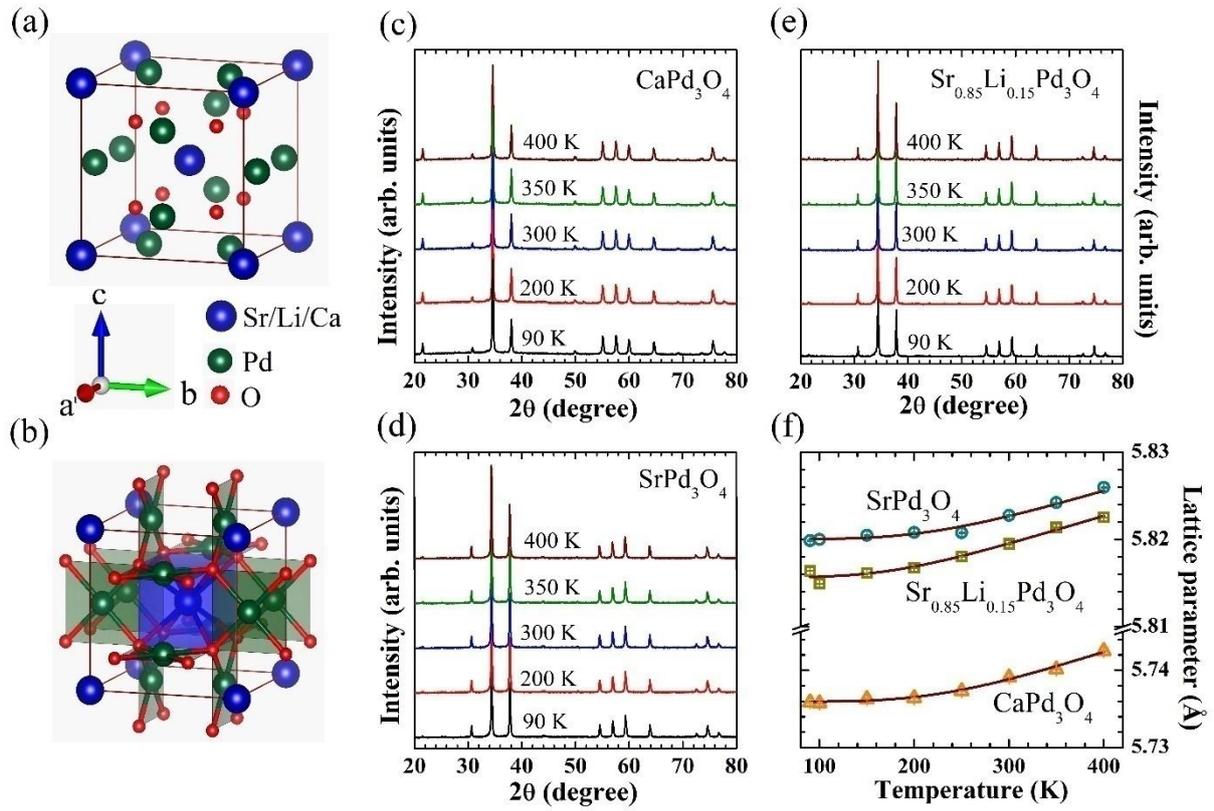

Figure 1: (a) Crystallographic unit cell, (b) showing AO$_8$ cube and PdO$_4$ square planes (c, d, e) X-ray diffraction pattern collected at a few temperatures and (f) lattice parameters as a function of temperature for APd$_3$O$_4$ with A = Ca, Sr, and Sr$_{0.85}$Li$_{0.15}$. Error bars in (f) are within the symbol size.

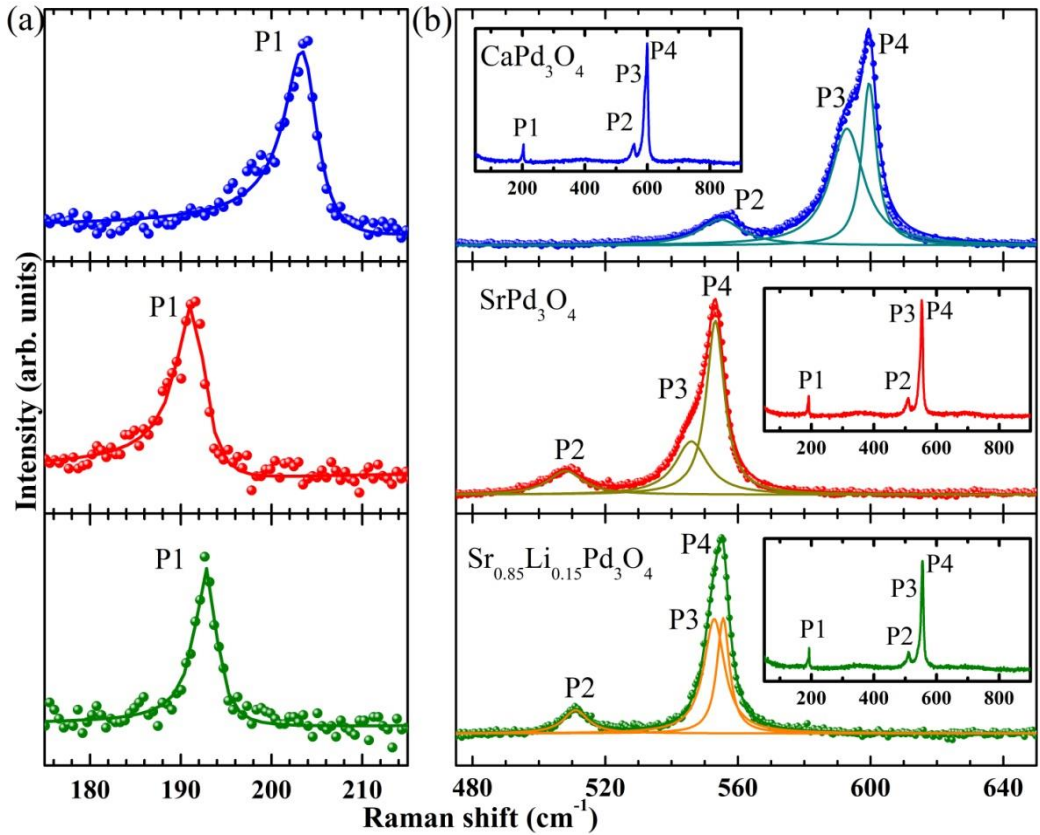

Figure 2: Raman spectra of APd$_3$O$_4$ with A = Ca, Sr, and Sr$_{0.85}$Li$_{0.15}$ collected at room temperature. (a) Enlarged view of P1 is shown for the clarity of spectral shape. The solid lines overlaying the data of P1 mode are Fano lineshape analysis. The solid lines in (b) are Lorentzian fit. Insets show the entire spectra.

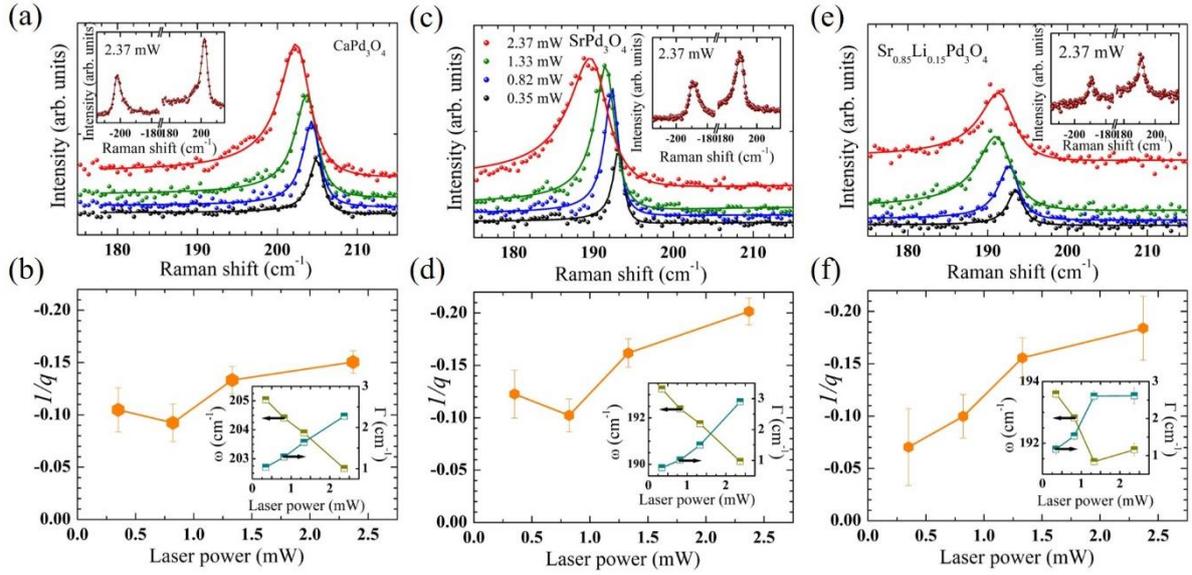

Figure 3: (a, c, e) The response of the Raman mode P1 collected at a few different laser powers, and (b, d, f) the dependence of Fano asymmetry parameter (1/q) as a function of laser power for APd$_3$O$_4$ with A = Ca, Sr, and Sr$_{0.85}$Li$_{0.15}$. The solid lines overlaying the data in (a, c, e) are Fano lineshape analysis. The solid lines in (b, d, f) and their insets are guide to eye. The insets in (a, c, e) show the Stoke vs anti-Stokes spectra at the respective laser powers exhibiting the Fano asymmetry.

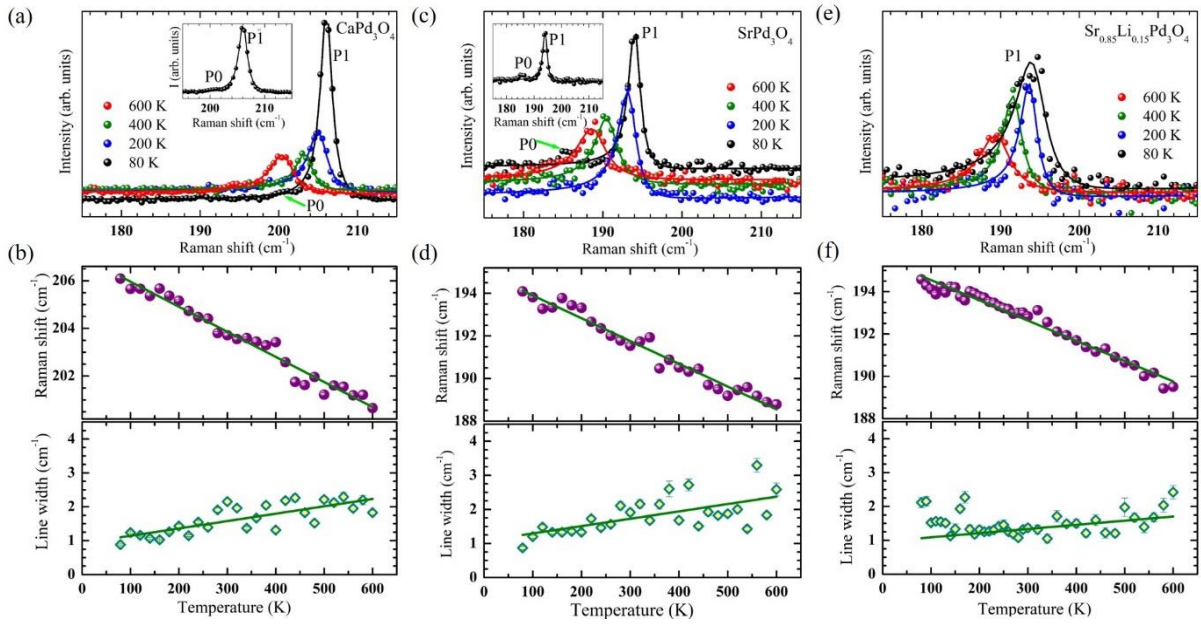

Figure 4: Raman spectra (shown for the mode P1) collected at a few temperatures (a, c, e); phonon frequency and linewidth as a function of temperature (b, d, f) for $APd_3O_4$ with A = Ca, Sr, and $Sr_{0.85}Li_{0.15}$. Green solid lines in (b,d,f) are fitting with anharmonic model (explained in the text). The insets in (a) and (c) show the presence of mode P0 at lowest measured temperature (80 K).

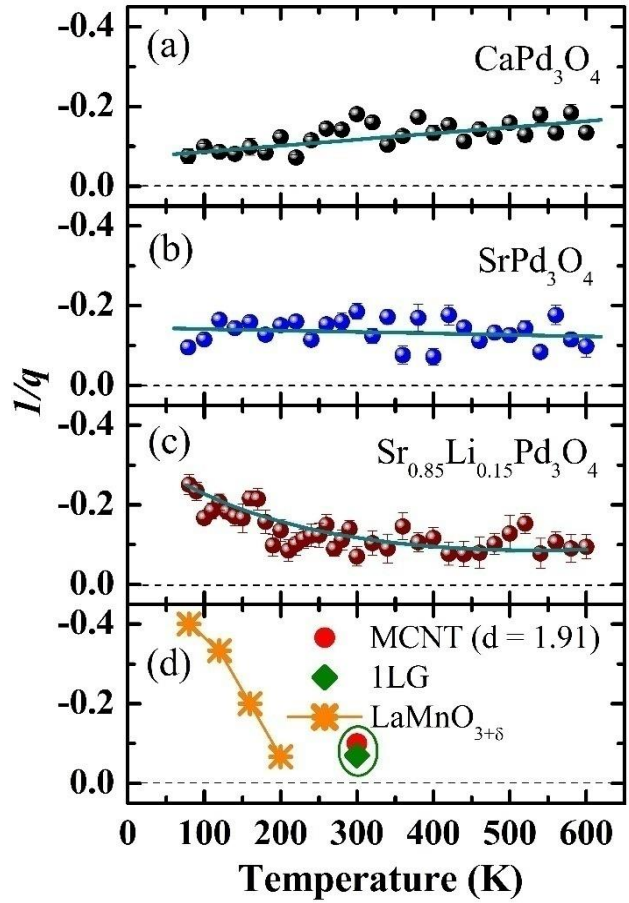

Figure 5: Fano asymmetry (1/q) as a function of temperature for $APd_3O_4$ with A = Ca, Sr, and $Sr_{0.85}Li_{0.15}$ (a,b,c), which are compared with the 1/q values reported (d) for a few known electron-phonon coupled systems (discussed in the text). Solid lines are guide to eye.

# Electron-phonon coupling in $APd_3O_4$: A = Sr, $Sr_{0.85}Li_{0.15}$ and Ca

Bommareddy Poojitha, B. H. Reddy, Aprajita Joshi, Ankit Kumar, Asif Ali,

R. S. Singh, and Surajit Saha*

*Department of Physics, Indian Institute of Science Education and Research, Bhopal 462066, India*

This supplementary material file contains additional information about the laser power- and temperature-dependent Raman spectra, and powder XRD. The figures and respective details are given below.

## Energy level diagram and electronic structure

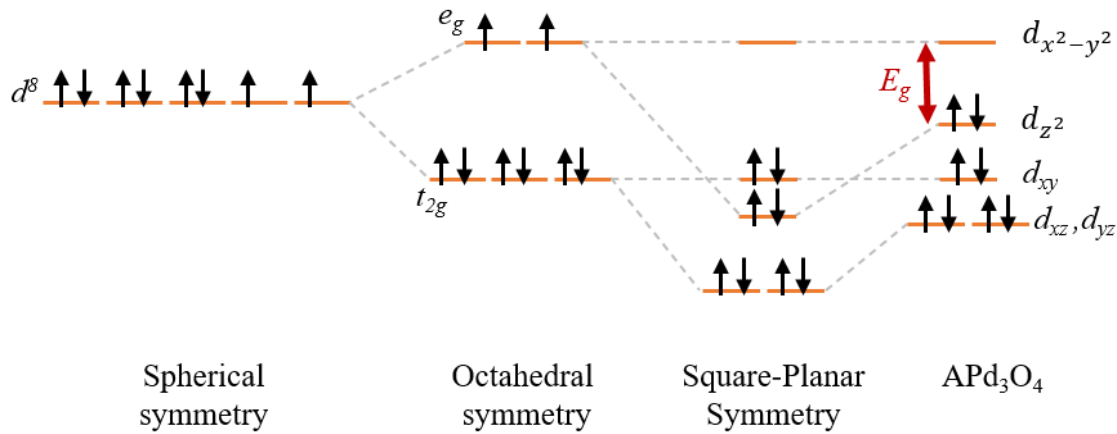

Figure S1: Crystal-field splitting for the general square planar case.

As shown in Figure S1, the octahedral crystal field splits d band into two $t_{2g}$ ($d_{xy}$, $d_{yz}$, $d_{xz}$) and $e_g$ ($d_{x^2-y^2}$, $d_{z^2}$) bands. While in the square planar symmetry, energy of $d_{xz}$, $d_{yz}$, and $d_{z^2}$ lowers in the absence of ligands along the z-axis and four ($d_{xz}$, $d_{yz}$), $d_{z^2}$, $d_{xy}$, and $d_{x^2-y^2}$ bands can be realized. This general picture is modified in case of APd$_3$O$_4$ where interaction between $d_{z^2}$ orbitals of stacked PdO$_4$ square planes lifts the $d_{xz}$, $d_{yz}$ and $d_{z^2}$ level. Here (Ca/Sr)Pd$_3$O$_4$ with $d^8$ electron configuration, have fully filled $d_{z^2}$ band and energy gap is observed between $d_{z^2}$ and $d_{x^2-y^2}$ bands [1].

Electronic structure has been calculated using FPLAPW method using Wien2k software within generalized gradient approximations (GGA) and GGA+U for various values of U. Figure S2 shows the band dispersion curves for CaPd$_3$O$_4$ (top panel) and SrPd$_3$O$_4$ (bottom panel) for both GGA (left) and GGA+U (right) along with Pd partial DOS corresponding to various orbitals. It is to note here that the GGA calculations do not exhibit a gap at the Fermi level in contrast to insulating/semiconducting transport. It is GGA+U calculations which opens up the gap for U > 4 eV. GGA+U calculations with U = 4.5 eV exhibit gaps of 0.215 eV for CaPd$_3$O$_4$ and 0.205 eV for SrPd$_3$O$_4$ commensurate with transport behaviour. These results suggest that the electron correlation plays an important role in these ternary palladates. Partial DOS shown in the right most panel for GGA+U calculations suggests that the gap appears between topmost filled $d_{z^2}$ band and completely unoccupied $d_{x^2-y^2}$ band.

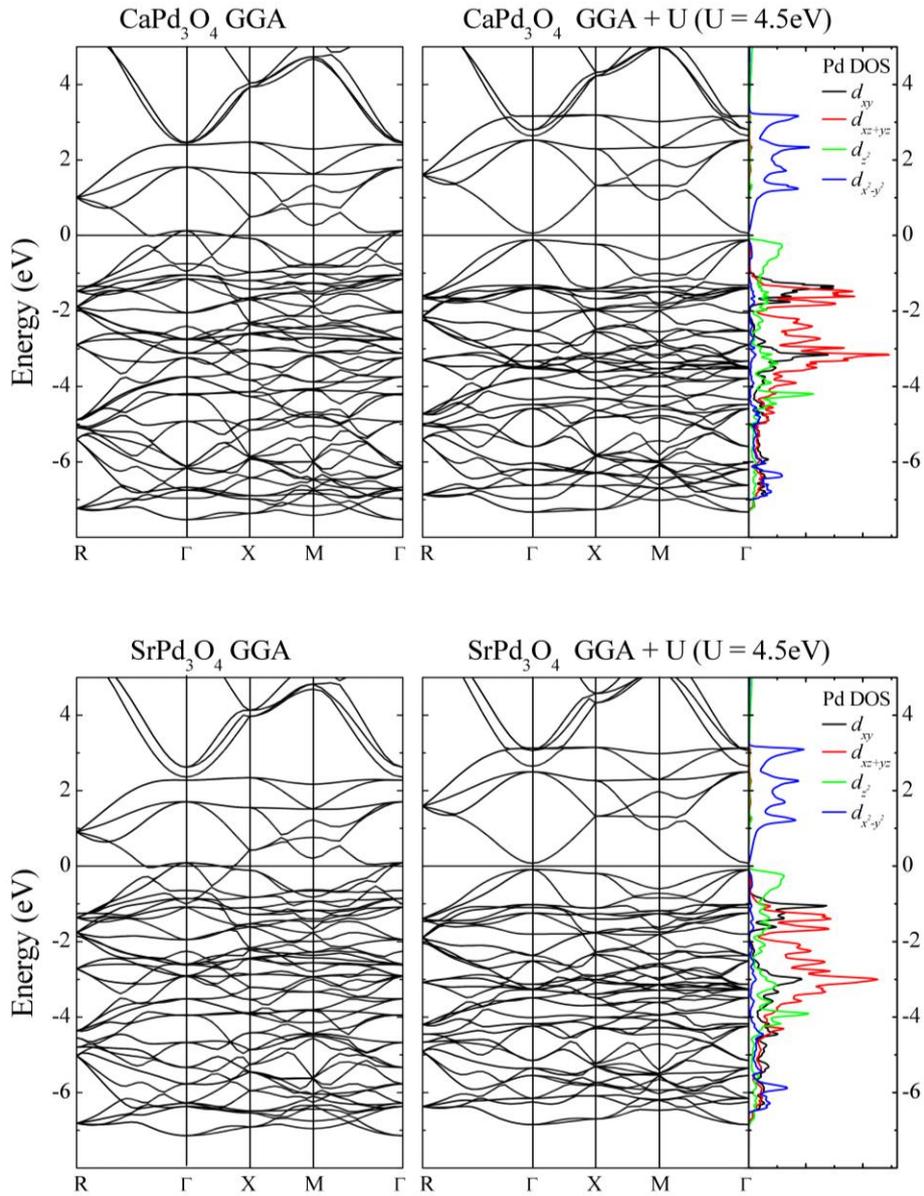

Figure S2: Band dispersion curves for $CaPd_3O_4$ (top panel) and $SrPd_3O_4$ (bottom panel) for both GGA (left) and GGA+U (right) along with Pd partial DOS corresponding to various orbitals.

**Justification for Fano line-shape of P1 mode**

Since the phonon calculations suggest two phonons having close energies around 190 cm$^{-1}$ (one at 177 and another at 193 cm$^{-1}$) we have verified fitting two Lorentzian peaks inside the profile of P1 mode and comparing with Fano analysis. It is very clear from the Figure S3 that the P1 shows a clear Fano line-shape justifying the electron-phonon coupling.

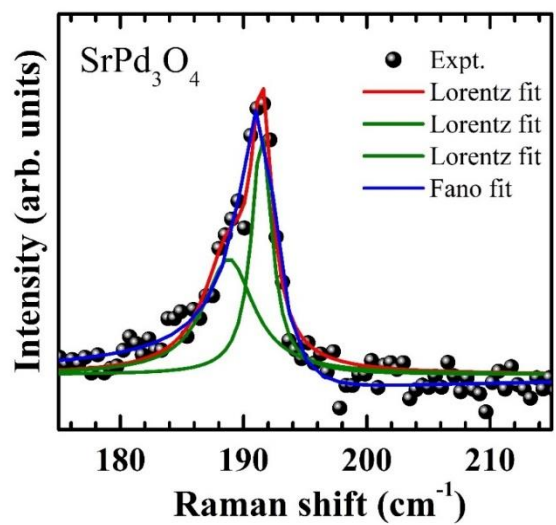

Figure S3: Fitting the experimental Raman line-shape with Fano as well as two-Lorentzian profiles. Fano fits the asymmetric profile better than Lorentzian taking care of the asymmetry on the low-energy side of the band, especially the higher baseline on the low-energy side.

**Laser power-dependence of Raman modes**

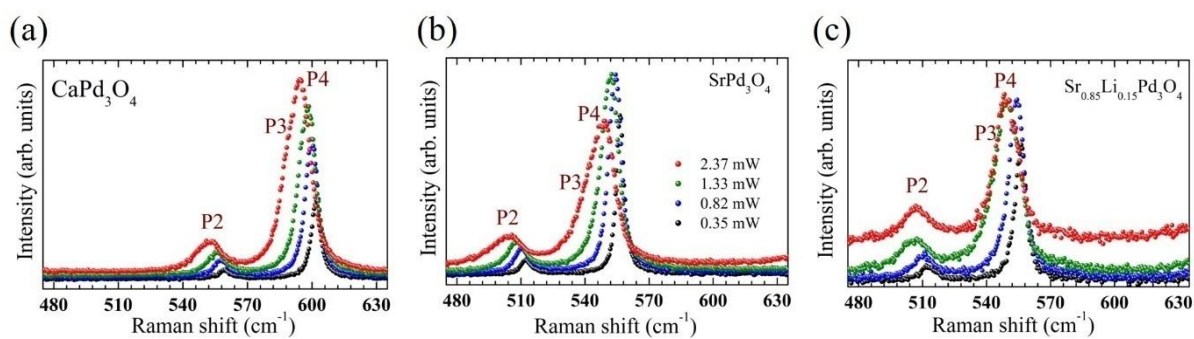

Figure S4. Raman spectra of $APd_3O_4$ (A = Ca, Sr, and $Sr_{0.85}Li_{0.15}$) collected at a few different laser powers.

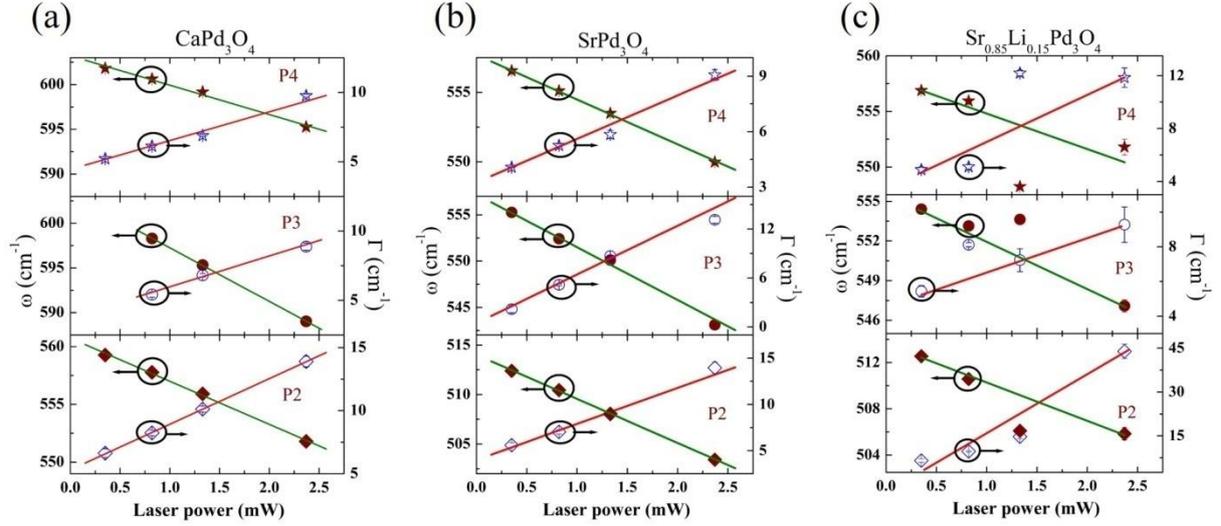

Figure S5. Phonon frequency and linewidth as a function of laser power for APd$_3$O$_4$ (A = Ca, Sr, and Sr$_{0.85}$Li$_{0.15}$). Solid lines are guide to eye.

The dependence of the additional modes (P2-P4) on the laser power can be seen in Figures S4 and S5. As can be seen in Figure S4, the modes show a red-shift with increasing laser-power (also depicted in Figure S4) with broadening (Figure S5). The changes in the modes with laser power can be attributed to the laser-power induced local heating, a behaviour that is analogous to temperature-induced effect due to phonon anharmonicity (discussed below).

**Temperature dependence of Raman modes**

Figure S6 shows the Raman spectra of APd$_3$O$_4$ (A = Sr, Sr$_{0.85}$Li$_{0.15}$, and Ca) collected at a few temperatures. The number of phonon modes remains unchanged throughout the investigated temperature range (80-600 K) which indicates the absence of structural phase transition as also evidenced by x-ray diffraction measurements discussed in the main text. The phonon parameters (frequency and linewidth) are extracted by Lorentzian fitting and plotted in Figures S7 and S8. The temperature-dependence of phonon frequency and linewidth are analysed by the anharmonic model. According to the model, the temperature-dependence of the phonon frequency due to cubic anharmonicity (three-phonon process) is given by [2, 3]:

$$\omega_{anh}(T) = \omega_0 + A\left[1 + \frac{2}{\left(e^{\frac{\hbar\omega_0}{2k_BT}}-1\right)}\right] \qquad (S1)$$

Similarly, the temperature-dependence of phonon linewidth due to cubic anharmonicity can be written as:

$$\Gamma_{anh}(T) = \Gamma_0 + C\left[1 + \frac{2}{\left(e^{\frac{\hbar\omega_0}{2k_B T}}-1\right)}\right] \qquad (S2)$$

where $\omega_0$ and $\Gamma_0$ are frequency and linewidth of the phonon at absolute zero temperature, *A,* and *C,* are cubic anharmonic coefficients for frequency and linewidth, respectively, $\hbar$ is reduced Planck constant, $k_B$ is Boltzmann constant and $T$ is the variable temperature.

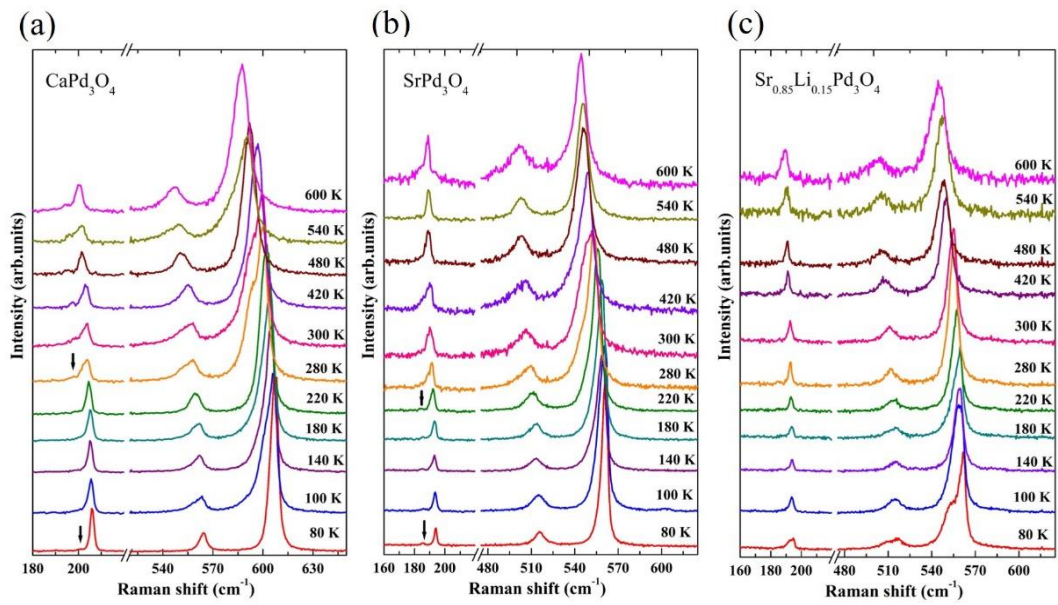

Figure S6. Raman spectra of APd$_3$O$_4$(A = Ca, Sr, and Sr$_{0.85}$Li$_{0.15}$) collected at a few temperatures. The mode P0 is marked with arrow.

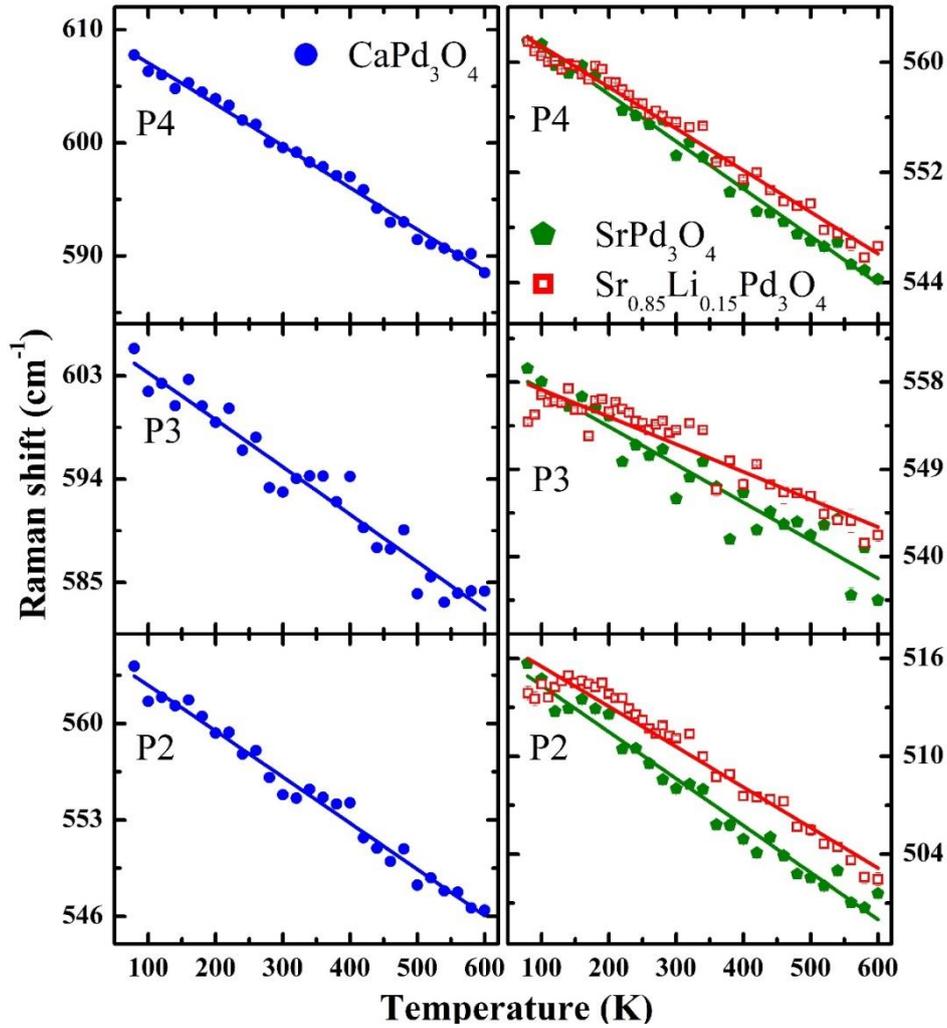

Figure S7. Frequency of phonons as a function of temperature for $APd_3O_4$ (A = Ca, Sr, and $Sr_{0.85}Li_{0.15}$). Solid lines represent anharmonic fitting (Eq. S1).

To conclude, the behaviour of the modes P2, P3, and P4 with temperature and laser power can be attributed to phonon anharmonicity showing normal thermal dependence whereas the mode P1 shows Fano asymmetric behaviour responding to the thermal changes (temperature and laser power) indicating the presence of its coupling with the electron degrees of freedom.

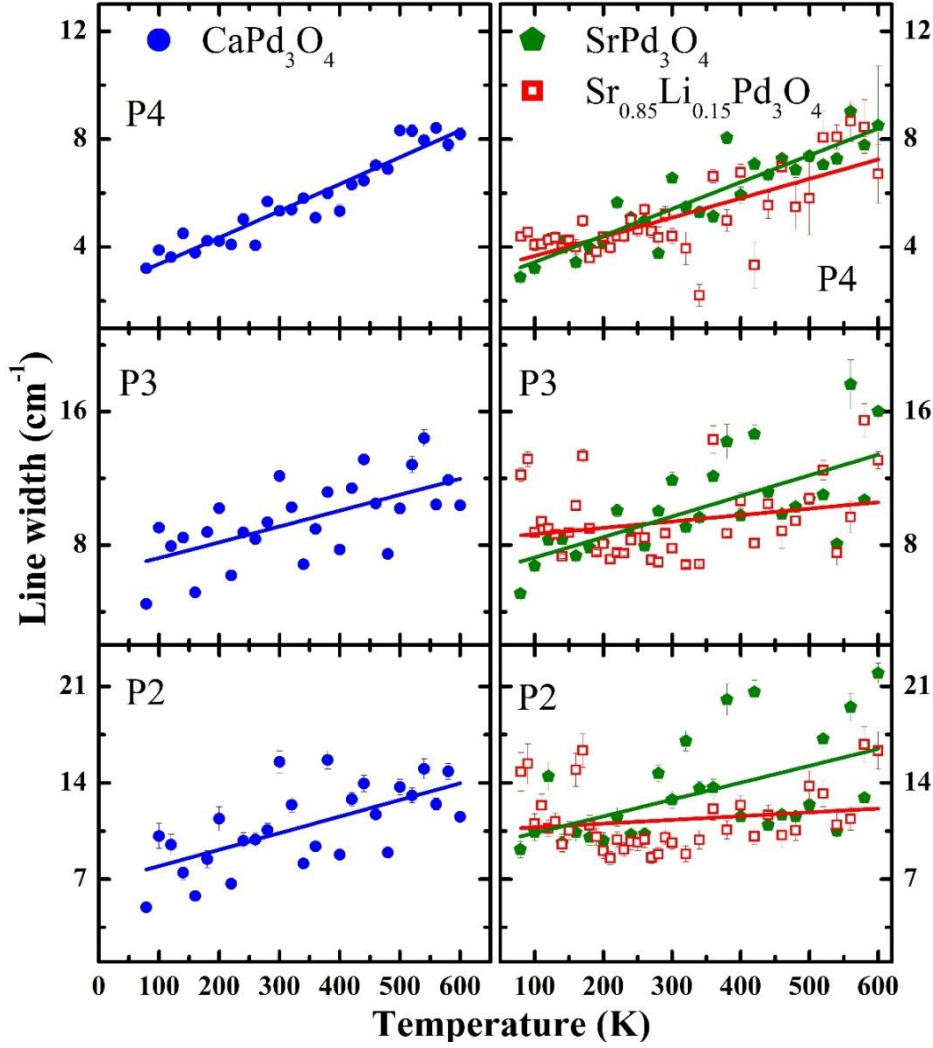

Figure S8. Linewidth of phonons as a function of temperature for $APd_3O_4$ (A = Ca, Sr, and $Sr_{0.85}Li_{0.15}$). Solid lines represent anharmonic fitting (Eq. S2).

Phonon calculation details

Our first-principle calculations within the density functional theory (DFT) for the Γ-point phonons of $SrPd_3O_4$ uses projector augmented wave method (PAW) implemented in Quantum Espresso. The exchange correlation used was Perdew-Burke-Ernzerhof (PBEsol). The calculation was performed systematically for which the step-by-step procedure is explained below.

Step 1: Lattice parameter optimization: The variable cell relaxation was done using cubic lattice ($i_{brav}$= 1), energy cut-off 150 Ry, and 4×4×4 monkhorst-Pack grid for optimization. This gives the lattice constant (11 Bohr unit) at which the total energy is minimum and structure is stable.

Step 2: The self-consistent field (scf) calculation was performed for minimizing total energy for different energy cut-off and k-grid mesh.

- $E_{cutoff}$ was varied from 100 Ry to 700 Ry by keeping k point fixed at 4×4×4 to get the $E_{cutoff}$ at which the total energy is minimum. The lowest $E_{cutoff}$ should always be more than provided in pseudopotential file.
- The scf calculation was repeated keeping $E_{cutoff}$ fixed this time (at which total energy was minimum) and k points were varied from 4×4×4 to 12×12×12 to obtain the k points information at which energy was minimum.

Step 3: Final scf calculation was performed to stabilize the structure by using k points grid and $E_{cutoff}$ obtained from Step 2 (values at minimum energy).

Step 4: Now, the phonon calculation was performed which uses data from output folder created by scf in step 3.

Step 5: The acoustic sum rules (ASR) is imposed after the phonon calculation by using the code *dynmat.x* which imposes the ASR on the elements of the dynamical matrix and diagonalizes it.

Table S1: Calculated phonon frequencies along with corresponding mode symmetries

| Phonon frequency | Symmetry |
|---|---|
| 0.00 | $T_{1u}$ |
| 150.62 | $T_{1u}$ |
| 154.07 | $A_{2g}$ |
| 157.48 | $T_{1g}$ |
| 161.22 | $T_{1u}$ |
| 163.32 | $T_{2u}$ |
| 177.09 | $E_g$ |
| 193.05 | $T_{2g}$ |
| 255.36 | $T_{2u}$ |
| 316.61 | $A_{2u}$ |
| 369.50 | $T_{1u}$ |
| 371.54 | $T_{1g}$ |
| 441.86 | $A_{2g}$ |
| 490.87 | $E_u$ |
| 495.68 | $T_{2u}$ |
| 512.52 | $T_{2g}$ |
| 531.54 | $T_{1g}$ |
| 539.67 | $E_g$ |
| 551.43 | $T_{1u}$ |

**Origin of electron-phonon coupling and its dependence on temperature and doping:**

To recall that $CaPd_3O_4$ and $SrPd_3O_4$ are narrow band-gap semiconductors with $E_g \sim 0.25$ eV and $\sim 0.22$ eV, respectively [3]. These band-gaps are small and hence the thermal energy is sufficient to excite finite number of carriers (electrons) into the conduction band at any finite temperature. These electrons may interact with the phonons in the material. On the other hand, $Sr_{0.85}Li_{0.15}Pd_3O_4$ is metallic in nature and, therefore, a finite number of electrons exists at the Fermi level at any temperature. Further, during Raman scattering measurements, we incident a laser beam of energy 2.33 eV (532 nm) which induces photo-excited carriers in these materials. Hence, both thermally- and photo-excited carriers contribute to the observed electron-phonon coupling in all the three compounds. Note that $CaPd_3O_4$ is slightly more insulating than $SrPd_3O_4$. In case of semiconductors/insulators, the intrinsic number carriers increase with increasing temperature. As a result, EPC increases with increasing temperature which can be seen in $CaPd_3O_4$. However, in even narrower band-gap semiconductors or metallic systems, as the temperature increases, the electron-electron scattering also becomes important which reduces the EPC, as we have seen in case of $SrPd_3O_4$ and $Sr_{0.85}Li_{0.15}Pd_3O_4$, and also as reported for $LaMnO_{3+\delta}$ (Figure 5 in the main text).

**References:**

1. Lamontagne, Leslie M. Schoop, and Ram Seshadri, Phys. Rev. B 99, 195148 (2019).
2. P.G. Klemens, Anharmonic decay of optical phonons, Phys. Rev. 148, 845 (1966).
3. M. Balkanski, R.F. Wallis, E. Haro, Anharmonic effects in light scattering due to optical phonons in silicon, Phys. Rev. B 28, 1928 (1983).
4. Leo K. Lamontagne, Geneva Laurita, Michael Knight, Huma Yusuf, Jerry Hu, Ram Seshadri, and Katharine Page, Inorg. Chem. 56, 5158 (2017).